\documentclass[fleqn,twoside]{article}
\usepackage{espcrc2}

\usepackage[dvips]{graphicx}
\usepackage[figuresright]{rotating}
\usepackage{epsf}
\usepackage{amssymb}

\newcommand{\AmS}{{\protect\the\textfont2
  A\kern-.1667em\lower.5ex\hbox{M}\kern-.125emS}}
\newcommand{\lsim}{\stackrel{<}{_\sim}}
\newcommand{\gsim}{\stackrel{>}{_\sim}}
\newcommand{\Frac}[2]{\frac{\displaystyle #1}{\displaystyle #2}}

\title{The Hadronic Cross--Section in the Resonance Energy Region
\thanks{Report IFIC/03-53. Talk given by J.~Portol\'es at the Workshop on 
Hadronic Cross--Section at Low Energy (SIGHAD03), 8th-10th October 2003,
Pisa (Italy).}}

\author{J. Portol\'es \address[Espanya]{Departament de F\'{\i}sica Te\`orica,
IFIC,
CSIC-Universitat de Val\`encia, \\ Apt. Correus 22085, E-46071 Val\`encia,
Spain}, 
        P.D. Ruiz--Femen\'{\i}a \addressmark[Espanya] }
       
\begin{document}

\begin{abstract}
We study the hadronic vacuum polarization in the resonance energy region, 
using the framework given by the Resonance Effective Theory of QCD. We consider 
the incorporation of vector--pseudoscalar meson loops that give, inclusively,
three and four pseudoscalar meson cuts. 
After resummation we achieve a QCD--based inclusive parameterization of the
correlator, hence of the hadronic cross--section in the energy region populated
by resonances.
\vspace{1pc}
\end{abstract}

\maketitle

\section{Introduction}
\hspace*{0.3cm}The hadronic spectrum from $e^+e^-$ annihilation in 
the energy range between 1 and 2 GeV  
exhibits a rather rich and complex structure.
Theoretically, the region $E\gsim M_{\rho}$ (with $M_{\rho}$ the mass of 
the $\rho(770)$ resonance), being far away from the chiral domain,
is poorly known due to the
intricate non-perturbative dynamics of QCD. The 
conventional approach to extract the hadronic matrix elements 
of the relevant QCD currents
has mainly relied on the available experimental information, such
as $e^+e^-\to hadrons$  or semileptonic decays. From these data,
the hadronic observables
have been obtained either by direct integration of the data
or by {\it ad hoc} parameterizations lousily inspired by QCD.
Both approaches have an obvious drawback: they do not tell
us much about the physics which lies behind. Even when we can
obviate
the physical interpretation, we shall keep in mind that fitting
(or integrating) procedures inherit all the uncertainties
associated to the experimental data, making it very 
difficult to define the accuracy of the results. An estimation of
the theoretical errors introduced by the above techniques
is always a matter of discussion. Recall, for
example, the running of the QED fine structure constant $\alpha(s)$ 
and the anomalous 
magnetic moment of the muon. These are observables whose 
theoretical predictions
are limited by loop effects from hadronic vacuum polarization. Both
magnitudes are related via dispersion relations to the hadronic 
production rate in $e^+e^-$ annihilation, which can be
evaluated using $e^+e¯$ data and hadronic $\tau$ decays. It is clear
that the apparent discrepancy between the measured value for 
the anomalous magnetic moment of the muon and the Standard Model 
prediction requires a careful review of the theoretical
uncertainties associated to the hadronic contribution
to accurately determine the size of this deviation.
An analysis of these observables in a model-independent way 
could clarify the issue. 
\par
Attempts based on effective
actions of QCD have achieved a remarkable success in describing
the data for energies up to 1 GeV, and indeed suggest that this approach may
be continued to higher energies. The pion vector form factor at very low
energies has been calculated in chiral perturbation theory, 
allowing to describe
the $e^+e^-\to \pi^+\pi^-$ data in this region very accurately with
the known values of the chiral parameters \cite{GL85}. Concerning
the muon anomalous magnetic moment, the use
of the chiral expansion for the two-pion contribution
at $E\le 0.5$ GeV 
has dramatically decreased its error as compared
to previous estimations directly obtained from the raw data.  
At higher energies ($E\sim M_{\rho}$), the appropriate framework to
implement QCD information is
Resonance Chiral Theory (R$\chi$T). This scheme has been the starting point
of several works devoted to the study of the pion 
form factor in the region close to the $\rho(770)$ mass
\cite{GP97,GP00,SP02},
which have also implemented features provided by the $1/N_C$ expansion,
resummation techniques 
and other important constraints such as analyticity and unitarity.
\par
Our goal is to provide a QCD--based parameterization of the hadronic
cross--section in the resonance driven 
1-2 GeV region. To proceed we will derive an expression
for the vector--vector current correlator 
following similar methods to those used in the works just mentioned.
With this aim, we outline here the general strategy to follow,
mainly focusing in the technical part of the analysis. As we shall see,
the practical implementation of our results shall require further
investigations. Among other interesting applications, this project
could cast some light on the above--mentioned issue of the anomalous
magnetic moment of the muon, for which about 90$\%
$ of the total hadronic contribution comes from the energy region $E\le2$ GeV.
\par
A more thorough explanation of the procedure put forward here is given
in Refs.~\cite{nos}.

\section{The Effective Action of QCD~: Resonance Chiral Theory}
\hspace*{0.3cm}The low--energy behaviour of QCD for the light quark
sector $(u,d,s)$ is known to be ruled by the spontaneous breaking of
chiral symmetry that set up the lightest hadron degrees of freedom,
identified with the octet of pseudoscalar mesons. The corresponding 
effective realization of QCD describing the interactions between the
Goldstone fields is called Chiral Perturbation Theory \cite{chpt}, and
its effective Lagrangian to lowest order in derivatives, ${\cal O}(p^2)$,
is given by~:
\begin{equation}
{\cal L}_{\chi}^{(2)}=\frac{F^2}{4}\langle u_{\mu}
u^{\mu} + \chi _+ \rangle \ ,
\label{eq:op2}
\end{equation}
where
\begin{eqnarray}
u_{\mu} & = & i [ u^{\dagger}(\partial_{\mu}-i r_{\mu})u-
u(\partial_{\mu}-i \ell_{\mu})u^{\dagger} ] \ , \nonumber \\ 
\chi_{\pm} & = & u^{\dagger}\chi u^{\dagger}\pm u\chi^{\dagger} u\ \ 
 , \ \
\chi=2B_0(s+ip) \, .
\end{eqnarray}
The unitary matrix in flavour space 
\begin{equation}
u(\phi)=\exp \left\{ i\frac{\Phi}{\sqrt{2}\,F} \right\} \; \; \; ,
\end{equation}
is a (non-linear) parameterization of the Goldstone octet of fields, 
identified with the mesons $\pi$, $K$ and $\eta$.
The external hermitian matrix fields $r_{\mu}$, $\ell_{\mu}$, $s$ and $p$ 
promote the global 
SU$(3)_{\mathrm{R}}\times$SU$(3)_{\mathrm{L}}$ symmetry 
of the Lagrangian to a local one, and generate Green functions of 
quark currents
by taking appropriate functional derivatives.
The ${\cal L}_{\chi}^{(2)}$ Lagrangian is settled by fixing the unknown
$F$ and $B_0$ parameters from the phenomenology~: 
$F \simeq F_{\pi} \simeq 92.4 \, \mbox{MeV}$
is the decay constant of the charged pion and 
$B_0 F^2 = - \langle 0 | \bar{\psi}\psi | 0 \rangle_0$ in the chiral limit.  
\par
Starting with the $\rho(770)$, the spectroscopy reveals the existence of 
multiple 
vector meson resonances participating in the relevant physics up to 
$E \sim 2 \, \mbox{GeV}$. These can be classified in $U(3)_{\mathrm{V}}$ 
nonets and must be included as explicit degrees of freedom in order to 
describe the hadron dynamics. In this work we will attach to the lightest
multiplet of vector resonances participating in the I=1 vector--vector
currents correlator. The generalization to several multiplets is rather
straightforward \cite{nos}.
\par
At the lowest order in derivatives, the chiral invariant Lagrangian
for the vector mesons and their interaction with Goldstone fields reads
\cite{rchit}, in the antisymmetric tensor formulation,
\begin{equation}
{\cal L}_{\mathrm{V}} \, =\,                  
{\cal L}_{\mathrm{K}}(V) 
+ {\cal L}_2(V)          
\, ,                                                                            
\label{eq:res_Lagrangian}                                                                            
\end{equation}
with kinetic terms
\begin{equation}
{\cal L}_{\mathrm{K}}(V)  =  
    - {1\over 2} \langle \nabla^\lambda V_{\lambda\mu}                      
\nabla_{\nu} V^{\nu\mu} -{M^2_V\over 2}  V_{\mu\nu} V^{\mu\nu}\rangle
 , 
\label{eq:kin_s}                                                                     
\end{equation}                                                                            
where $M_V$ is the mass of the lowest nonet of vector resonances
under U$(3)_{\mathrm{V}}$, and 
the covariant derivative 
\begin{eqnarray}
\nabla_{\mu}V & = & \partial_{\mu}V+[\Gamma_{\mu},V] , \\  
\Gamma_{\mu} & = & \frac{1}{2}\{ u^{\dagger}(\partial_{\mu} - i r_{\mu})u+
u(\partial_{\mu} - i \ell_{\mu})u^{\dagger}\, \} \,  , \nonumber
\end{eqnarray}
is defined in such a way that $\nabla_{\mu}V$ also transforms as a 
nonet under
the action of the group.                
For the interaction Lagrangian ${\cal L}_2(V)$ we have                
\begin{eqnarray} \label{eq:V_int}                                                                           
{\cal L}_2(V) & = &   {F_V\over 2\sqrt{2}} \,
\langle V_{\mu\nu} f_+^{\mu\nu}\rangle +                                   
{iG_V\over \sqrt{2}} \, \langle V_{\mu\nu} u^\mu u^\nu\rangle \ , \nonumber
\\
f_{\pm}^{\mu\nu} & = & u F_L^{\mu\nu}u^{\dagger}\pm u^{\dagger} 
F_R^{\mu\nu} u \, ,                                                                                                                                                   
\end{eqnarray}   
with $F_{L,R}^{\mu\nu}$ the field strength tensors of the left and right
external sources $\ell_{\mu}$ and $r_{\mu}$, and $F_V$, $G_V$ are real 
couplings. 
\par
The chiral couplings contained in ${\cal L}_2(V)$ 
only concern the even--intrinsic--parity sector. 
In Ref.~\cite{EG89a} it was shown that, up to ${\cal O}(p^4)$ in the
chiral counting, the effective Lagrangian
${\cal L}_{\chi V}\equiv{\cal L}_{\chi}^{(2)}+{\cal L}_{\mathrm{V}}$ 
is enough to satisfy the
short-distance QCD constraints where vector resonances play a significant
role. 
\par
Contributions of one--loop two--point diagrams involving a vector 
and a pseudoscalar mesons provide an inclusive description of exclusive 
channels with four (4$\pi$) or three ($\overline{K} K \pi$) pseudoscalars.
The relevant vertices violate intrinsic--parity and are given by the
resonance Lagrangian for the odd--intrinsic--parity sector which reads~:
\begin{equation}\label{eq:Lano}
{\cal L}_V^{\mathrm{odd}} =
 \sum_{a=1}^{7} \frac{c_a}{M_{V}} {\cal O}^a_{\mbox{\tiny{VJP}}}
 \, + \, \sum_{a=1}^{4} d_a {\cal O}^a_{\mbox{\tiny{VVP}}}
\ .\; 
\end{equation}
The new operators ${\cal O}_{\mbox{\tiny VJP}}^i$ and 
${\cal O}_{\mbox{\tiny VVP}}^i$ have been given explicitly in 
Ref.~\cite{noi}. The $c_a$ and $d_a$ real couplings, that 
are not fixed by the underlying symmetry properties are, in 
principle, unknown.
The set defined above is a complete basis for constructing 
vertices with only one-pseudoscalar; for a larger number of pseudoscalars
additional operators may emerge.  
\par
In the following we will consider the Effective Action of QCD given
by the R$\chi$T Lagrangian~:
\begin{equation}
{\cal L}_{\mathrm{ R \chi T}} \, = \, {\cal L}_{\chi}^{(2)} \, + \, 
{\cal L}_{\mathrm{V}} \, + \, {\cal L}_V^{\mathrm{odd}}\, .
\end{equation} 

\section{The vector--vector currents correlator~: Dyson-Schwinger 
resummation}
\hspace*{0.3cm}The main object of study in this work is the two-point 
function built from the
I=1 part of the electromagnetic current,  
\begin{eqnarray}
\Pi_{\mu\nu}^{33}(q^2)&=&i\int d^4x\,e^{iqx}
\langle 0|\,T[V^3_{\mu}(x)V^3_{\nu}(0)]\, |0\rangle
\nonumber\\[3mm]
&=&(q_{\mu}q_{\nu}-q^2g_{\mu\nu}) \ \Pi^{33}(q^2)\,,
\label{eq:VVdef}
\end{eqnarray}
with the vector current given by  
\begin{equation}
V^3_{\mu}=\frac{\delta S^{\chi}_R}{\delta v_3^{\mu}}\,,
\label{eq:V3def}
\end{equation}
being $S^{\chi}_R$ the effective action associated to 
${\cal L}_{\mathrm{ R \chi T}}$, and
the external vector field $v_{\mu}\equiv\frac{\lambda^a}{2}v_{\mu}^a$.
Current conservation has been used to extract the tensor structure of
the correlator in Eq.~(\ref{eq:VVdef}).         
The observable quantity we shall derive from the $\Pi^{33}(q^2)$ correlator
is the inclusive hadronic cross--section in the I=1 channel:
\begin{eqnarray}
R^{\mathrm{I}=1}_{had} & = & \frac{\sigma^{\mathrm{I}=1}(e^+e^-\to hadrons)}
{\sigma(e^+e^-\to\mu^+\mu^-)} \nonumber \\[3mm]
& = & 12\,\pi\,{\mbox{Im}}\,\Pi^{33}(q^2)\,.
\end{eqnarray}
At the one--loop level we consider two different types of absorptive terms
that contribute to the imaginary part of the correlator~: loops with two
pseudoscalars, arising from ${\cal L}_2(V)$ and ${\cal L}_{\chi}^{(2)}$,
and loops with one internal resonance, given by ${\cal L}_{R}^{\mathrm{odd}}$;
both can be attached to a vector meson or directly to the $V_{\mu}^3$ 
currents to build up the correlator. 
\par
We are not interested though in performing the evaluation of the
correlator up to one--loop only. The bare resonances acquire a finite 
width through resummation of quantum loops in perturbation theory. These
effects are subleading in the $1/N_C$ counting but must be accounted for
to avoid the singularities arising at energies close to the bare pole
of resonance propagators.
In order to obtain the dynamics in full
we would like to resummate all the possible contributions constructed 
in terms of the one--loop terms explained above. This we do now in turn.

\subsection{Two--pseudoscalar meson loops}
\begin{figure}[!htb]
\hspace*{-0.3cm} 
\includegraphics*[scale=0.51,clip]{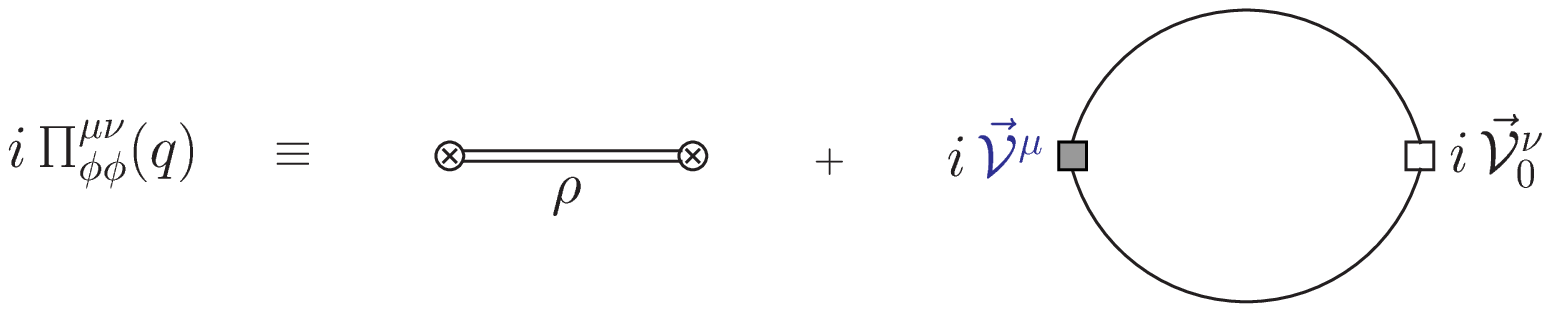}
\vspace*{-0.7cm}
\caption{\label{fig:f1}The $\Pi_{\phi \phi}^{\mu \nu}$ vector--vector currents
correlator with resummed pseudoscalar loops. Single lines stand for 
pseudoscalar mesons, double lines stand for vector resonances. }
\vspace*{-0.5cm}
\end{figure}

\begin{figure}[!htb]
\vspace*{-0.1cm}
\hspace*{-0.2cm} 
\includegraphics*[scale=0.50,clip]{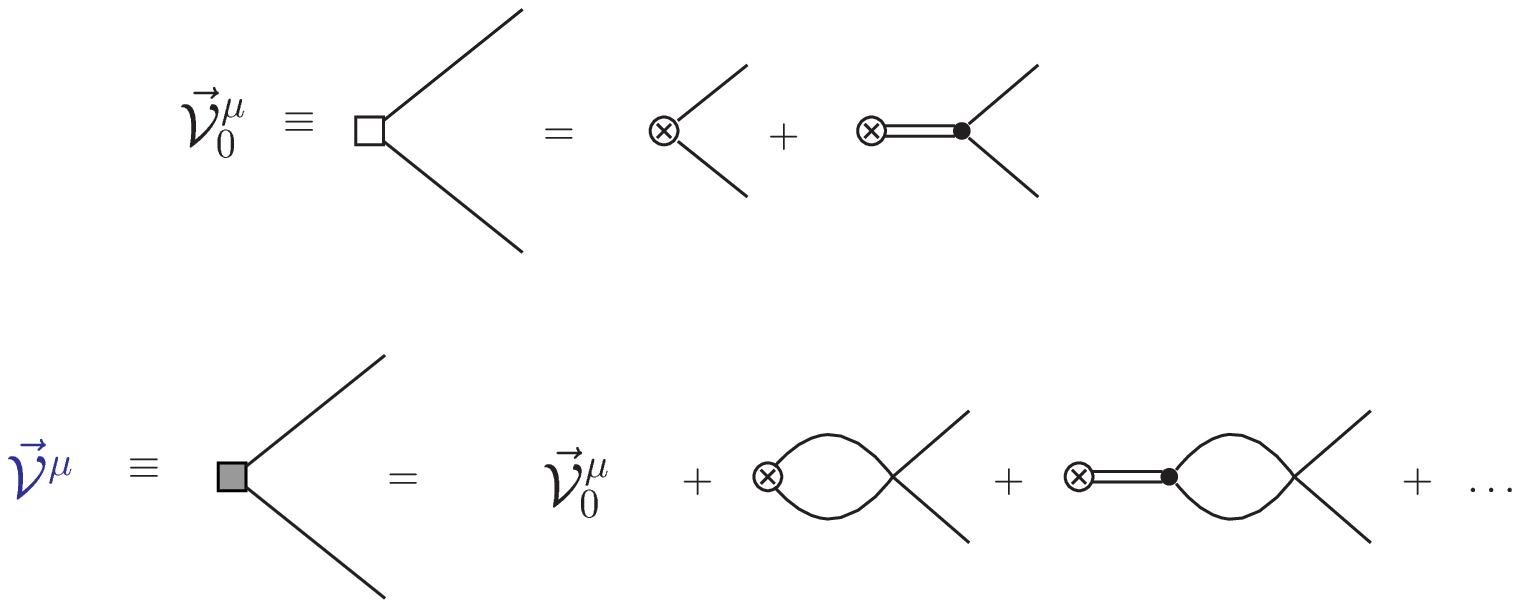}
\vspace*{-0.7cm}
\caption{\label{fig:f2} Definition of the off--shell effective current
vertices appearing in the resummation $\Pi_{\phi \phi}^{\mu \nu}$ in 
Fig.~\ref{fig:f1}. }
\vspace*{-0.5cm}
\end{figure}

\hspace*{0.3cm} These contributions have already been taken into account
in detail in Refs.~\cite{GP00,SP02}. The procedure is sketched in 
Fig.~\ref{fig:f1} where the vertices, generated by the 
vector form factor of the pseudoscalar mesons, are given in 
Fig.~\ref{fig:f2}.

The resummed two-point function finally reads~:
\begin{eqnarray}
\Pi_{\phi \phi}(q^2) & = & \frac{-4\left( 1 + \frac{F_{V} G_{V}}{F^2} 
 \frac{q^2}{M_{V}^2-q^2}\right)^2 \overline{B}_{22}}
{ 1 + \Big(1+\frac{2G_{V}^2}{F^2}\frac{q^2}{M_V^2-q^2}\Big)
\frac{2q^2}{F^2}\overline{B}_{22}}  \nonumber
\\[3mm]
& & +\frac{F_V^2}{M_V^2-q^2} ,
\label{eq:Pires}
\end{eqnarray}
where
$\overline{B}_{22}\equiv B_{22}[q^2,m_{\pi}^2,m_{\pi}^2]+\frac{1}{2}
B_{22}[q^2,m_{K}^2,m_{K}^2]$ and $B_{22}[q^2,m^2,m^2]$ is
the Passarino--Veltman two--point integral as given in 
Refs.~\cite{GP00,nos}.

\subsection{Vector--pseudoscalar meson loops}
\hspace*{0.3cm} These contributions arise from the odd--intrinsic--parity
couplings in ${\cal L}_V^{\mathrm{odd}}$. 
Here we limit ourselves to the easier case of one multiplet of vector mesons
and refer the reader to Ref.~\cite{nos} for a more complete treatment.
\par
\begin{figure}[!htb]
\hspace*{-0.1cm} 
\includegraphics*[scale=0.52,clip]{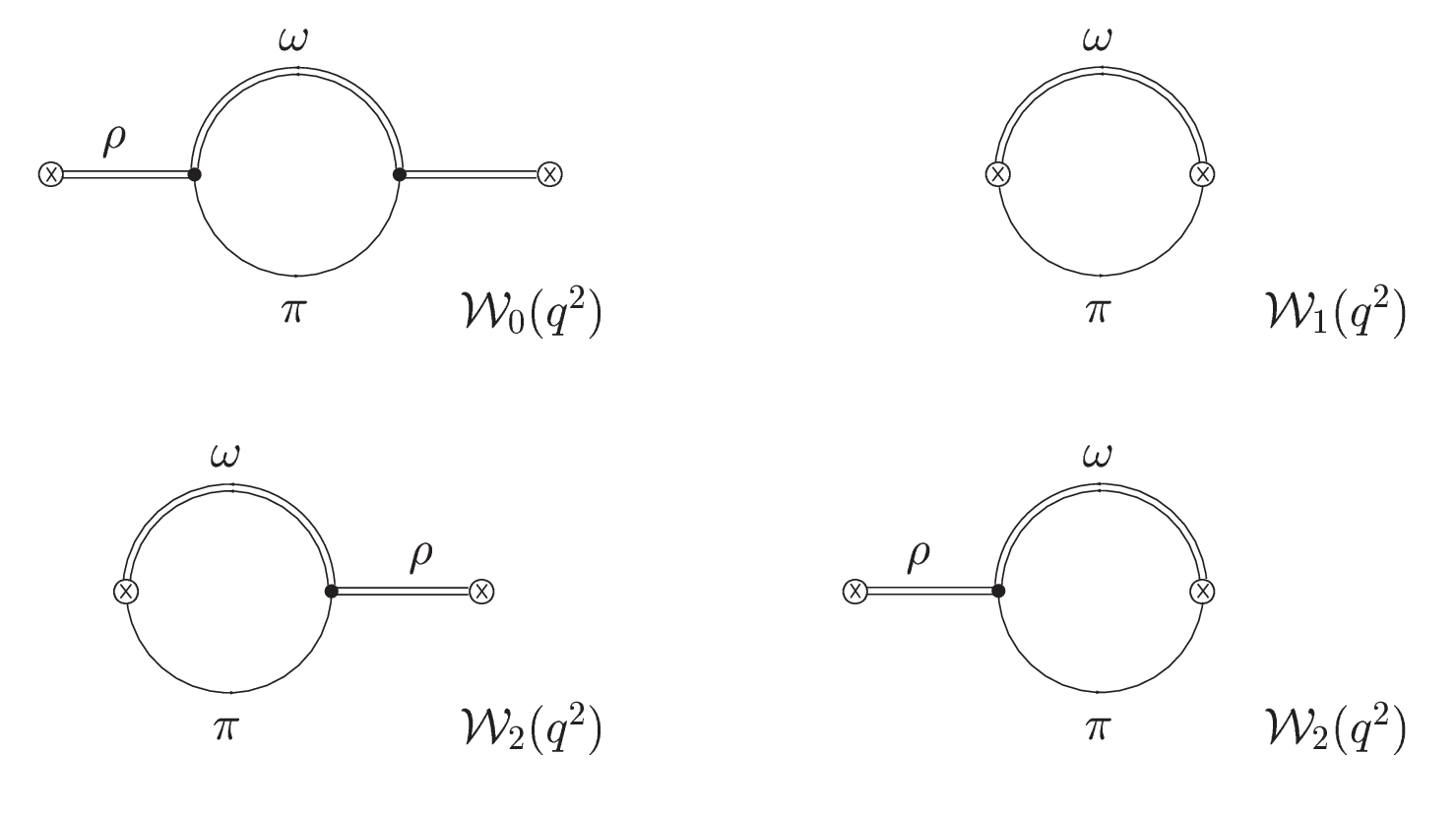}
\vspace*{-0.7cm}
\caption{\label{fig:f3} The vector--pseudoscalar mesons contribution
to the vector--vector correlator at one--loop. ${\cal W}_0$, 
${\cal W}_1$ and ${\cal W}_2$ are the invariant functions associated
to the loops according to Eq.~(\protect{\ref{eq:1loopCorrelator}}). }
\vspace*{-0.5cm}
\end{figure}

First of all we consider the one--loop contributions. The four allowed
topologies are shown in Fig.~\ref{fig:f3} and
the result reads~: 
\begin{eqnarray} \label{eq:1loopCorrelator}
\Pi_{V\phi}^{\mathrm{1-\ell oop}}(q^2) & = & \! \!  - \! \! \! \sum_{P=\pi,K}
\frac{C_P^2}{F^2}
  \Bigg\{ 
\frac{F_{V}^2 \, {\cal W}_{0,P}(q^2) }{(M_{V}^2-q^2)^2}
\nonumber \\
& & \\
&& \! \! \! \! +{\cal W}_{1,P}(q^2)
+4\,\frac{F_{V} \, {\cal W }_{2,P}(q^2)}{(M_{V}^2-q^2)}
\, \Bigg\}\,,\nonumber
\end{eqnarray}
where the constants $C_P$ are Clebsch--Gordan coefficients depending
on the pseudoscalar meson $P = \pi^0, K^+, K^-, K^0, \bar{K}^0$, 
running inside the loop (together with $\omega$ or $K^*$).

\subsubsection{The one--loop functions}
\hspace*{0.3cm}
The functions ${\cal W}_i(q^2)$ are divergent quantities which need
to be regularized. The full expression for these functions, obtained
following the $\overline{MS}$ subtraction scheme, can be found
in Ref.~\cite{nos}, though the renormalization program of R$\chi$T
remains an unexplored issue.
\par
Alternatively, we can bypass the lack of a consistent renormalization
procedure by using a dispersion technique to regularize the real
part of the ${\cal W}_i(q^2)$ functions from their well--defined
imaginary parts~: 
\begin{eqnarray} \label{eq:dispe}
{\cal W}_{i,P}(s)& = & \sum_{k=0}^{N_i}a_i^{(k)}\,s^k \\[3mm]
& & +\frac{s^{N_i+1}}{\pi}
\int_{s_{th}}^{\infty}
ds^{\prime}\frac{\mbox{Im}\,{\cal W}_{i,P}(s^{\prime})}{s^{\prime (N_i+1)}
(s^{\prime}-s)}\,, \nonumber
\end{eqnarray}
where $s_{th} = (M_V + m_P)^2$. The number of subtraction constants  
needed depends
on the behaviour of the spectral densities Im ${\cal W}_i(q^2)$ at 
large $q^2$ and our evaluation shows that $N_0 = 3$, $N_1=1$ and $N_2 = 2$.

\subsubsection{The coupling constants of ${\cal L}_V^{\mathrm{odd}}$}
\hspace*{0.3cm} The one--loop functions ${\cal W}_i(q^2)$ depend
on the coupling constants $c_a$ and $d_a$ of
${\cal L}_V^{\mathrm{odd}}$ in Eq.~(\ref{eq:Lano}). In Ref.~\cite{noi}
we have put forward a procedure to obtain information on those couplings
from QCD itself. This exploits the fact that the QCD Green's function
$\langle V_{\mu} V_{\nu} P \rangle$ of vector ($V_{\mu}$) and pseudoscalar
($P$) QCD currents is an order parameter of the spontaneous chiral symmetry
breaking of QCD, hence it does not get perturbative contributions in the
chiral limit. Accordingly we match the evaluation of the three--point
function in the R$\chi$T framework, at leading order in the $1/N_C$ 
expansion, with the first OPE coefficient of the Green's function within
QCD. As a result we fix 5 relations between the unknown $c_a$ and $d_a$
couplings. We have also shown that these results agree well with the
phenomenology of odd--intrinsic--parity violating processes.
\par
Quite remarkably the combinations of those couplings appearing in
$\mbox{Im} \, {\cal W}_i$ get fixed by the short--distance conditions 
extracted, using this method, in Ref.~\cite{noi}. This is an important
result of our work because, taking a look at the dispersion relation
description in Eq.~(\ref{eq:dispe}), we realize that the 
analytic structure of the functions ${\cal W}_i$ is completely fixed
except for a polynomial whose coefficients are the subtraction constants
encoding our lack of knowledge on the renormalization procedure.

\subsubsection{Resummation of the one--loop contributions}
\hspace*{0.3cm} The inclusion into $\Pi_{\phi \phi}(q^2)$ in 
Eq.~(\ref{eq:Pires}) of the resummation driven by the one--loop
diagrams in Fig.~\ref{fig:f3} is rather straightforward if one
notices that the role of the latter is to modify structures already
given by the two--pseudoscalar meson resummation.
\par
\begin{figure}[!htb]
\hspace*{-0.1cm} 
\includegraphics*[scale=0.52,clip]{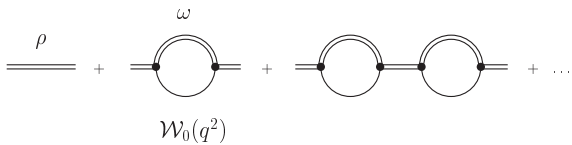}
\vspace*{-0.7cm}
\caption{\label{fig:f4} The vector meson propagator with $\omega$-$\pi$
and $K^*$-$K$ insertions. }
\vspace*{-0.5cm}
\end{figure}
The procedure reduces to 
consider the following steps~:
\begin{itemize}
\item[i)] As shown in Fig.~\ref{fig:f4} the vector--pseudoscalar meson
loops modify the propagator of the vector meson. The final effect is
to generate a shift of the position of the corresponding pole~:
\begin{equation}
M_V^2 \longrightarrow M_V^2 \, + \, {\cal W}_0(q^2) \, ,
\end{equation}
where 
\begin{equation}
{\cal W}_{0}(q^2)=\sum_{P=\pi,K}\frac{C_P^2}{F^2}\,{\cal W}_{0,P}(q^2) \; 
\end{equation}
\begin{figure}[!htb]
\hspace*{0.1cm} 
\includegraphics*[scale=0.56,clip]{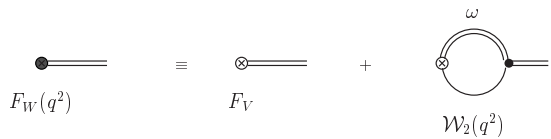}
\vspace*{-0.7cm}
\caption{\label{fig:f5} The vertex loop correction to $F_V$. }
\vspace*{-0.5cm}
\end{figure}
\item[ii)] The ${\cal W}_{2,P}$ functions introduce a $q^2$ dependence
on the coupling of the vector meson to the external current 
(Fig.~\ref{fig:f5})~:
\begin{equation}
F_V  \longrightarrow {F_W}(q^2) \equiv F_V + {\cal W}_2(q^2) \; ,
\end{equation} with
\begin{equation}
{\cal W}_2(q^2) =   2\sqrt{2} \!  \sum_{P=\pi,K}\frac{C_P^2}{F^2}
{\cal W}_{2,P}(q^2)\,.
\end{equation}
\end{itemize}
Including both corrections we finally obtain an analytical expression
for the $\Pi^{33}(q^2)$ two--point function which accounts for both
the two--pseudoscalar loops and the vector--pseudoscalar meson loops
that provide the two--particle absorptive cuts emerged from the
odd--intrinsic--parity sector~:
\begin{equation}\label{eq:Pires2}
 \Pi^{33}(q^2) = 
\end{equation}
\begin{displaymath}
 \Frac{-4\left( 1 + \frac{F_{W}(q^2) G_{V}}{F^2} 
 \frac{q^2}{M_{V}^2-q^2+{\cal W}_{0}}\right)^2 \overline{B}_{22}}
{ 1 + \Big(1+\frac{2G_{V}^2}{F^2}\frac{q^2}{M_V^2-q^2+{\cal W}_{0}}
\Big)
\frac{2q^2}{F^2}\overline{B}_{22}} 
\end{displaymath}
\begin{displaymath}
 +\Frac{F_W^2(q^2)}{M_V^2-q^2+{\cal W}_{0}}
+\sum_{P=\pi,K}\frac{C_P^2}{F^2}\,{\cal W}_{1,P}(q^2)\,. 
\end{displaymath}

\section{Summary}
\hspace*{0.3cm} The expression for $\Pi^{33}(q^2)$ in Eq.~(\ref{eq:Pires2}) is
the main result of this study. It provides a QCD--based parameterization
of the inclusive hadronic cross--section carrying information on the most
relevant exclusive final states of two, three and four pseudoscalar mesons
in the isovector channel when only one multiplet of vector mesons is
considered.
\par
It is clear that a more complete description of this observable in the 
whole resonance region ($M_{\rho} \lsim E \lsim 2 \, \mbox{GeV}$) requires
the inclusion of heavier multiplets of vector resonances. A look to the
RPP \cite{pdg} shows the existence of three of those multiplets in the 
$I=1$ channel commanded by $\rho(770)$, $\rho(1450)$ and 
$\rho(1700)$. Accordingly, a full description of the vector--vector 
correlator and the hadronic cross--section needs to consider this
circumstance that has been implemented and carried out in detail 
(for N multiplets)
in Ref.~\cite{nos}. However, in this case, we are not at the point of
providing reasonable predictability due to our lack of knowledge on the
increasing number of coupling constants that appear as new multiplets are 
introduced in the R$\chi$T.
\par
Nevertheless the procedure deserves a close analysis. When only one
multiplet is included, as we have done here, we have managed to obtain
information on the couplings from QCD itself (at leading order in 
the $1/N_C$ expansion) \cite{noi} and we have seen
that the short--distance conditions fix unambiguously the contributions,
but for some polynomials which depend 
on the regularization procedure in R$\chi$T. Consequently it is 
compulsory to carry the non--trivial study of the implementation of QCD 
constraints when more than one multiplet are present in order to 
complement the resummation procedure accomplished here.
\par
Though further investigations are called for, our study has shown that 
the use of effective theories of QCD in the intermediate energy region, 
populated by resonances, provides a powerful tool to endow the basic
information of the underlying theory into the hadron phenomenology
in an essentially model--independent way. 
\vspace*{0.9cm} \\
\noindent
{\bf Acknowledgements}\par
\vspace{0.2cm}
\noindent 
J.~Portol\'es wishes to thank to the organizers of the SIGHAD03 Workshop
for an excellent and very interesting meeting. 
This work has been supported in part by TMR EURIDICE, EC Contract No. 
HPRN-CT-2002-00311, by MCYT (Spain) under grant FPA2001-3031, and
by ERDF funds from the European Commission.

\end{document}